\documentclass[pra,twocolumn,amsmath,amssymb,superscriptaddress]{revtex4-1}
\usepackage{epsfig,amsmath}
\usepackage{subfigure}
\usepackage{graphicx}
\usepackage{dcolumn}
\usepackage{stmaryrd}
\usepackage{mathrsfs}
\usepackage{pifont}
\usepackage{amsthm}
\usepackage{amssymb}
\usepackage{bm}
\usepackage{latexsym}
\usepackage[colorlinks=true,linkcolor=blue,citecolor=blue]{hyperref}
\usepackage{color}
\usepackage{epstopdf}

\usepackage{booktabs}
\usepackage{threeparttable}
\usepackage{multirow}

\begin{document}

\title{Universal quantum control over Majorana zero modes}

\author{Zhu-yao Jin}
\affiliation{School of Physics, Zhejiang University, Hangzhou 310027, Zhejiang, China}

\author{Jun Jing}
\email{Contact author: jingjun@zju.edu.cn}
\affiliation{School of Physics, Zhejiang University, Hangzhou 310027, Zhejiang, China}

\date{\today}

\begin{abstract}
Majorana zero mode (MZM) exhibits inherent resilience to local parametric fluctuations, due to the topological protection mechanism in the non-Abelian braiding statistics of the anyonic quasiparticles. In this paper, we construct the braiding operations between an arbitrary pair in three MZMs under the theoretical framework of universal quantum control. Largely detuned driving fields on the mediator, a local defect of lattice, enable indirect and tunable exchange interaction between arbitrary two MZMs. The braiding operations can then emerge through the time evolution along the universal nonadiabatic passages, whose robustness against driving-field errors and quasiparticle poisoning can be substantially enhanced by the rapid modulation over the passage-dependent global phase. Moreover, a chiral transfer for population on MZMs along the universal passage can be perfectly demonstrated in both clockwise and counterclockwise manners without eliminating the mediator defect. Our protocol is linear scalable and provides an avenue towards the universal quantum control over MZMs, which is fundamental and essential for topological quantum computation.
\end{abstract}

\maketitle

\section{Introduction}

Certain problems~\cite{Sergey2018Quantum} can show provable advantages of quantum computer over its classical counterpart, i.e., Shor's algorithm for factoring large numbers~\cite{Calderbank1996Good,Ekert1996Quantum,Kivlichan2020Improved} and Grover's algorithm for amplifying probability of target states~\cite{Grover1997,Nagib2025}. However, practical quantum computer demands large-scale quantum systems with extensive gate operations~\cite{Gidney2019HowTF}. Each elementary operation can introduce systematic errors that accumulate exponentially, posing significance of error correction schemes, such as the Shor's code~\cite{Shor1995Scheme}. That scheme encodes a single logical qubit into nine physical qubits and can detect and correct single-qubit errors. Another paradigm for error correction is topological quantum computation~\cite{Nayak2008NonAbelian,Michael1998PNP,Kitaev2001Unpaired,Kitaev2003Fault,
Das2005Topologically,Stern2006Proposed,Bonderson2006Detecting}, in which qubits are encoded in the non-Abelian anyonic quasiparticles~\cite{Nayak2008NonAbelian}. Quantum information encoded in those nonlocal degrees of freedom is believed to maintain intrinsic fault tolerance against local parametric fluctuations.

Anyons fundamentally bifurcate into the Abelian~\cite{Wilczek1999Fractional} and non-Abelian categories~\cite{Read2000Paired,Ivanov2001NonAbelian,Alicea2011NonAbelian,Sarma2015Majorana}. The unitary evolutions resulting from braiding Abelian anyons yield the accumulation of phases: the wave function of multiple Abelian anyons acquires a fractional phase factor $e^{i\theta}$ with $\theta\neq0,\pi$ on the exchange of quasiparticles. Bosons ($\theta=0$) and fermions ($\theta=\pi$) are their limiting cases~\cite{Wilczek1999Fractional}. Besides the phase accumulation, the braid group~\cite{Ivanov2001NonAbelian,Alicea2011NonAbelian,Sarma2015Majorana} associated with the non-Abelian anyons enables nontrivial unitary operations among the degenerate many-quasiparticle states. Due to the fact that the qubits encoded in the global degenerate states are robust against the local systematic errors~\cite{Witzel2006Quantum}, the non-Abelian anyons can serve as the primitive units of topological quantum computation.

Majorana zero mode (MZM), first theoretically predicted in the Kitaev chain model~\cite{Kitaev2001Unpaired}, represents the most elementary realization of the non-Abelian anyons. The Kitaev chain models a one-dimensional system of spinless fermions and was once regarded as a somewhat unphysical toy model~\cite{Elliott2015Colloquium}. Nevertheless, the spinless fermions can be simulated through experimentally accessible proxy systems, such as the spin-orbit-coupling electron gases under Zeeman field, as suggested by the subsequent theoretical studies~\cite{Elliott2015Colloquium,Fu2008Superconducting,
Sau2010Generic,Alicea2010Majorana,Lutchyn2010Majorana,Oreg2010Helical,Nadj2013Proposal,Pientka2013Topological,Gao2024Majorana}. Experimental research for MZMs has been conducted in multiple physical platforms, including the superconducting systems~\cite{Stevan2014Observation,Howon2018Toward,Jiao2018Chiral,
Schneider2020Controlling,Jie2021Spin} and hybrid magnetic-superconductor systems~\cite{Schneider2021Topological,Schneider2022Precursors,Soldini2023Twodimensional}. Recently, indirect evidences for MZMs have been witnessed in the hybrid semiconductor-superconductor system~\cite{Microsoft2025Interferometric}, exhibiting a zero-bias conductance peak~\cite{Das2012Zerobias,Deng2012Anomalous,Finck2013Anomalous,Churchill2013Superconductor} and the topological protection against the variations of local parameters~\cite{Jiao2018Chiral,Jie2021Spin,Microsoft2025Interferometric}. The braiding of MZMs, however, remains unimplemented as the gold-standard signature for definitively establishing their non-Abelian statistics~\cite{Dvir2023Realization}.

Conventional braiding operations of MZMs were proposed under the adiabatic condition~\cite{Ivanov2001NonAbelian,Stone2006Fusion,Read2009NonAbelian,Sau2011Controlling,vanHeck2012Coulomb,Kraus2013Braiding} and governed by the Berry-Wilczek-Zee holonomy~\cite{Wilczek1984Appearance,Cheng2011Nonadiabatic}. In practice, the implementation of topological quantum computation based on the adiabatic braiding operations is constrained by the extended operation time. First, the prolonged exposure to the environment can induce ruinable decoherence effect on the quantum systems~\cite{Carmichael1999statistical}. Second, while the exact adiabatic solution is established within an infinite period of time, the realistic operation with a finite clock speed unavoidably entails nonzero nonadiabatic errors~\cite{Nayak2008NonAbelian}. Such errors will give rise to the leakage from the subspace of degenerate ground states to the external Hilbert space. Third, the long operation time of topological quantum computation increases the possibility of the quasiparticle poisoning~\cite{Flsensberg2010Tunneling,Catelani2011Quasiparticle,Leijnse2011Scheme,
Rainis2012Majorana,Karzig2021Quasiparticle,Liu2024Quasiparticle}. The quantum information encoded in the braiding process of MZMs will leak out of the topological subspace due to the excitation of the extra quasiparticle, which can break the parity conservation of the MZMs subspace. Although the measurement-based tools~\cite{Knapp2016Nature} have been used to mitigate the unwanted transition errors during adiabatic braid operations~\cite{Bonderson2008Measurement,Bomantara2020Measurement}, a high-fidelity, robust, and fast braiding operation remains desired for MZMs.

Inspired by transitionless quantum driving~\cite{Berry2009Transition}, the shortcuts to non-Abelian braiding~\cite{Karzig2015Shortcuts,Wang2025Shortcuts} exploit the time-dependent and direct exchange interaction between MZMs as the counterdiabatic driving fields. But practical implementation of such interactions in semiconductor-superconductor systems is fundamentally difficult~\cite{Amundsen2024Colloquium}. In analog to nonadiabatic holonomic quantum computation~\cite{Sjoqvist2012Nonadiabatic,Liu2019Plug}, the nonadiabatic braiding operations~\cite{Yu2025Nonadiabatic} have been proposed for a two-MZM system mediated by the lattice defect (LD). However, the braiding operations under the parallel-transport condition~\cite{Jing2017Non} is sensitive to the systematic errors in the local driving fields on LD.

We here construct the braiding operations between any pair of three MZMs coupled to a mediator LD under local driving fields by the universal quantum control (UQC) theory~\cite{Jin2025Universal,Jin2025Entangling,Jin2025ErrCorr,Jin2025Rydberg,
Jin2025UniversalNon}. It is a theoretical framework that unifies the shortcuts to adiabaticity and holonomic quantum transformation and is beyond the parallel-transport condition. The degree of freedom of LD can be eliminated~\cite{Jin2025Rydberg,James2007Effective} when the driving fields are largely detuned; and then the indirect exchange interaction among MZMs can be established. Consequently, the braiding operations of arbitrary two MZMs can be performed via the universal passages of the system. Assisted by the fast-modulated global phases, our braiding operations become insensitive to the significant fluctuations in both transition frequency of LD and Rabi frequency of driving fields and to the quasiparticle poisoning. In addition, the clockwise and counterclockwise population transfers of three MZMs can be demonstrated by the time evolution along the passage that is superposed by LD excited state and MZMs. Our protocol is also directly extendable when the mediator LD coupled to more MZMs.

The remainder of this paper is structured as follows. In Sec.~\ref{modelHamiltonian}, we introduce a general model that is consisted of three Kitaev chains coupled to the same lattice defect and the full Hamiltonian for LD and MZMs. In Sec.~\ref{BraidId}, we derive the effective Hamiltonian with the largely detuned driving fields on LD, based on which the braiding operations can be established via universal nonadiabatic passages. In Sec~\ref{Correct}, we employ a rapidly varying global phase to correct systematic errors in driving fields, which gives rise to fault-tolerant, fast, and high-fidelity braiding operations for MZMs. In Sec.~\ref{quasipoison}, we evaluate our error-correction mechanism under the quasiparticle poisoning and discuss a combination strategy with the measurement-based braiding protocol. Section~\ref{chiral} is devoted to designing the chiral population transfer of three MZMs in the presence of LD excitation. The whole work is concluded in Sec.~\ref{conclusion}. Appendix~\ref{UniPara} presents the UQC theory in the parametric space with an ideal setting, followed by the error-correction extension in Appendix~\ref{UniParaErr}.

\section{Theoretical framework}\label{Braid}

\subsection{Model and Hamiltonian}\label{modelHamiltonian}

\begin{figure}[htbp]
\centering
\includegraphics[width=0.8\linewidth]{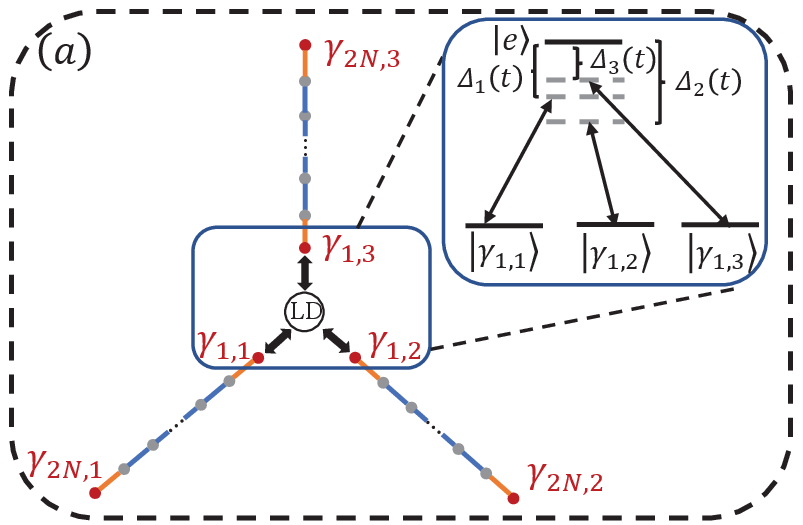}
\includegraphics[width=0.8\linewidth]{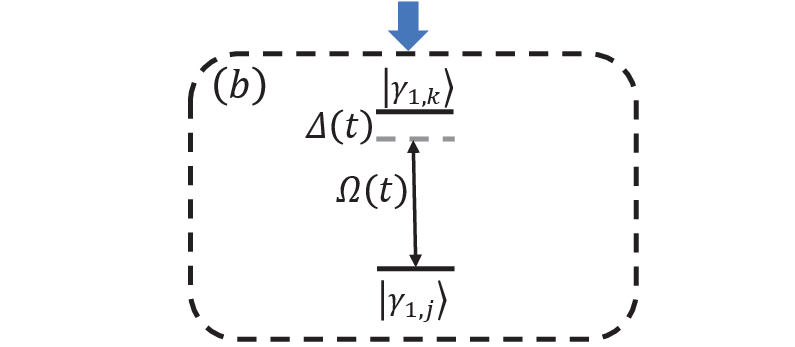}
\caption{(a) Sketch of an atomic system comprising a lattice defect and three one-dimensional $N$-site Kitaev chains. LD can be coupled to MZMs $\gamma_{1,1}$, $\gamma_{1,2}$, and $\gamma_{1,3}$ through the local off-resonant driving fields. Inset: the four-level transition diagram for the effective subspace of LD and MZMs. (b) Effective two-level transition diagram for the braiding operation between two MZMs $\gamma_{1,j}$ and $\gamma_{1,k}$, $j\ne k\in{1,2,3}$.}\label{model}
\end{figure}

Our braiding protocol is conducted on the atomic-chain system as shown in Fig.~\ref{model}(a). The system comprises a two-level system (lattice defect) as a mediator, consisting of an excited state $|e\rangle$ and a ground state $|g\rangle$, and three one-dimensional Kitaev chains~\cite{Kitaev2001Unpaired,Kitaev2003Fault} of open boundary condition, each of which is composed of $N$ sites. The system Hamiltonian can be written as~\cite{Kitaev2001Unpaired,Kitaev2003Fault} ($\hbar\equiv1$)
\begin{equation}\label{Ham}
\begin{aligned}
H(t)&=\sum_{k=1}^3\Big[-\mu\sum_{n=1}^Nc_{n,k}^\dagger c_{n,k}-\sum_{n=1}^{N-1}(Jc_{n+1,k}^\dagger c_{n,k}\\
&+gc_{n+1,k}^\dagger c_{n,k}^\dagger+{\rm H.c.})\Big]+\omega(t)\sigma_+\sigma_-,
\end{aligned}
\end{equation}
where $\sigma_+\equiv|e\rangle\langle g|$ ($\sigma_-=|g\rangle\langle e|$) and $c_{n,k}^\dagger$ ($c_{n,k}$) are the creation (annihilation) operators for LD and the $n$th site of the $k$th Kitaev chain, respectively. The fermions in the Kitaev chains are assumed to have the same transition frequency $\mu$. $\omega(t)$ is the transition frequency of the two-level lattice defect, which is set as a constant in our protocol despite it can be tuned via an external magnetic field in solid-state spin system~\cite{Wolfowicz2021Quantum}. $J$ and $g$ are the hopping and pairing coupling strength between neighboring fermions in the chains, respectively.

The Hamiltonian~(\ref{Ham}) admits an alternative representation in terms of Majorana fermion quasiparticles, which emerge as bounded pairs in the system. Specifically, the Majorana fermions of the $k$th Kitaev chain, $1\le k\le 3$, can be defined as~\cite{Kitaev2001Unpaired,Kitaev2003Fault}
\begin{equation}\label{DefMajorana}
\gamma_{2n-1,k}=c_{n,k}+c_{n,k}^\dagger, \quad \gamma_{2n,k}=-i(c_{n,k}-c_{n,k}^\dagger),\\
\end{equation}
where $n$ runs from $1$ to $N$. As defined in Eq.~(\ref{DefMajorana}), Majorana fermions are characterized by two fundamental properties: self-conjugation, $\gamma_{n,k}^\dagger=\gamma_{n,k}$ and anticommutation relations $\{\gamma_{n,j}, \gamma_{m,k}\}=2\delta_{nm}\delta_{jk}$, where $n$ and $m$ run from $1$ to $2N$.

In the current hybrid semiconductor-superconductor systems~\cite{Sau2012Realizing,Dvir2023Realization,Shi2024Probing,Bordin2025Enhanced,
Nitsch2025Poor,Microsoft2025Interferometric}, both LD and the fermions in the Kitaev chains are constituted by the quantum dots (QDs). The hopping coupling $J$ and pairing coupling $g$ between the neighboring quantum dots can be attained by elastic-to-tunneling and crossed Andreev reflection, respectively, mediated by a superconducting nanowire which hosts an Andreev bound state~\cite{Liu2022Tunable,Bordin2023Tunable}. Experimentally~\cite{Sau2012Realizing,
Dvir2023Realization,Shi2024Probing,Bordin2025Enhanced,Nitsch2025Poor,Microsoft2025Interferometric}, both the coupling strengths $J$ and $g$ and the transition frequencies of LD $\omega$ and QDs $\mu$, can be individually manipulated by the gate voltages applied to the superconducting segments. $\omega/2\pi$ and $\mu/2\pi$ can be tuned within the range of $\sim(-10,10)$ GHz. Using the parametric setting of $\mu=0$ and $J=-g$~\cite{Kitaev2001Unpaired,Kitaev2003Fault} and Eq.~(\ref{DefMajorana}), the Hamiltonian~(\ref{Ham}) is transformed to be
\begin{equation}\label{HamMaj}
H(t)=ig\sum_{k=1}^3\sum_{n=1}^{N-1}\gamma_{2n+1,k}\gamma_{2n,k}+\omega\sigma_+\sigma_-.
\end{equation}
It indicates that two decoupled Majorana zero modes appear at both ends of each Kitaev chain, i.e., $\gamma_{1,k}$ and $\gamma_{2N,k}$.

We desire to construct the nonadiabatic braiding operations between the chosen pair of MZMs $\gamma_{1,j}$ and $\gamma_{1,k}$, $j\neq k$, via a mediator LD. When the system is initially in the vacuum state $|0\rangle$, its dynamics is constrained in the subspace spanned by the states $|\gamma_{1,k}\rangle=\gamma_{1,k}|0\rangle$ in the absence of LD. In addition to the free Hamiltonian~(\ref{HamMaj}), we introduce the local time-dependent hopping interaction $J_k(t)$ and pairing interaction $g_k(t)$ between LD and the end fermion of three Kitaev chains~\cite{Dvir2023Realization,Shi2024Probing,Bordin2025Enhanced,
Microsoft2025Interferometric,Nitsch2025Poor}. Then the full Hamiltonian can be expressed as~\cite{Yu2025Nonadiabatic}
\begin{equation}\label{HamDri}
H_{\rm tot}(t)=H(t)+H_d(t),
\end{equation}
where the interaction or driving Hamiltonian reads,
\begin{equation}\label{Hdrive}
H_d(t)=\sum_{k=1}^3\left[J_k(t)\sigma_+c_{1,k}+g_k(t)\sigma_+c_{1,k}^\dagger+{\rm H.c.}\right].
\end{equation}
Similar to $J$ and $g$, the coupling strengths $J_k(t)$ and $g_k(t)$ can be effectively realized by the adiabatic elimination of the superconducting circuit in hybrid semiconductor-superconductor systems~\cite{Dvir2023Realization,Shi2024Probing,Bordin2025Enhanced,
Microsoft2025Interferometric,Nitsch2025Poor}. In experiments, they are determined by the tunneling amplitudes for spin-conserving and spin-flipping processes and the BCS coherence factors and the eigenenergy of the Andreev bound state formulated in the superconductor circuit. Therefore, both $J_k(t)$ and $g_k(t)$ can become fully tunable via the gate voltages applied to the superconducting segments. Here it is required that $J_k(t)=g_k(t)=2\Omega_k(t)\cos[\int_0^t\omega_k(s)ds+\varphi_k]$. In typical experimental conditions, $J_k/2\pi$ and $g_k/2\pi$ can be tuned within the range of $\sim(-1,1)$ GHz~\cite{Dvir2023Realization,Shi2024Probing,Bordin2025Enhanced,Microsoft2025Interferometric,
Nitsch2025Poor}. Consequently, the Rabi frequencies are about $\Omega_k(t)/2\pi\sim(-0.5,0.5)$ GHz. The Hamiltonian~(\ref{Hdrive}) has been extensively studied, both theoretically~\cite{Liu2013Floquet,Li2014Tunable,Long2021Nonadiabatic,
Nathan2021Quasiperiodic,Xu2023Dynamics} and experimentally~\cite{Cooper2019Topological,Sebastian2025Braiding}, in cold atoms and quantum-dot arrays.

Adapting from MZMs, the full Hamiltonian $H_{\rm rot}(t)$~(\ref{HamDri}) suggests that there are no interactions between LD and the rest Majorana fermions $\gamma_{2n}$ and $\gamma_{2n+1}$, $1\le n\le N-1$. Then we can focus on
\begin{equation}\label{HamDriRedu}
\begin{aligned}
H_{\rm tot}(t)&=\omega\sigma_+\sigma_-+\Big\{2\sum_{k=1}^3\Omega_k(t)\\
&\times\cos\left[\int_0^t\omega_k(s)ds+\varphi_k\right]|e\rangle\langle\gamma_{1,k}|+{\rm H.c.}\Big\}.
\end{aligned}
\end{equation}
In the rotating frame with respect to $H_0=\omega\sigma_+\sigma_-$, it can be rewritten as a time-dependent control Hamiltonian
\begin{equation}\label{HamMrot}
\begin{aligned}
H_I(t)&=\sum_{k=1}^3\Omega_k(t)\Big\{e^{i\int_0^t[\omega_k(s)+\omega]ds+i\varphi_k}\\
&+e^{-i\int_0^t\Delta_k(s)ds-i\varphi_k}\Big\}|e\rangle\langle\gamma_{1,k}|+{\rm H.c.},
\end{aligned}
\end{equation}
where the detuning $\Delta_k(t)\equiv\omega_k(t)-\omega$. The relevant diagram of level transition can be found in the inset of Fig.~\ref{model}(a). The two exponential terms in Eq.~(\ref{HamMrot}) have the same physical resource, i.e., the time-dependent driving frequencies $\omega_k(t)$. By tuning the magnitudes of $\Delta_k(t)$, the transition between the states $|\gamma_{1,j}\rangle$ and $|\gamma_{1,k}\rangle$ can be effectively engineered as the following subsection.

\subsection{Braiding operations via UQC passage}\label{BraidId}

In this subsection, we construct the braiding operations between an arbitrary pair in three MZMs, i.e., $\gamma_{1,j}$ and $\gamma_{1,k}$, $j\ne k$, through a synthesis of the effective Hamiltonian~\cite{Jin2025Rydberg,James2007Effective} and UQC theory~\cite{Jin2025Universal,Jin2025Entangling,Jin2025ErrCorr,Jin2025Rydberg} (see Appendix~\ref{UniPara} for details). When the mediator LD is under the largely detuned driving fields, the system dynamics is governed by the effective Hamiltonian~\cite{Jin2025Rydberg,James2007Effective}
\begin{equation}\label{Heff}
H_{\rm eff}(t)=-\frac{i}{2}\left[H_I(t), \int_0^tH_I(s)ds\right].
\end{equation}
For simplicity, the Rabi frequencies in the control Hamiltonian~(\ref{HamMrot}) are set to be isotropic, i.e., $\Omega_1(t)=\Omega_2(t)=\Omega_3(t)$. The detunings are parameterized as $\Delta_j(t)=\Delta_0-\Delta(t)/2$ and $\Delta_k(t)=\Delta_0+\Delta(t)/2$ with the scaling detunings $\Delta_0$ and $\Delta(t)$. Then under the large detuning condition $\Delta_j(t),\Delta_k(t),|\Delta_j(t)-\Delta_l(t)|,|\Delta_k(t)-\Delta_l(t)|\gg\Omega_1(t)$ with $l\neq j,k \in\{1,2,3\}$ and the near-resonant condition $\Delta_k(t)-\Delta_j(t)=\Delta(t)\sim\Omega_1(t)$, the effective Hamiltonian~(\ref{Heff}) results in a two-body interaction Hamiltonian:
\begin{equation}\label{HameffMZ}
H_{\rm eff}(t)=\Omega(t)e^{i\int_0^t\Delta(s)ds+i\phi_{jk}}|\gamma_{1,j}\rangle\langle\gamma_{1,k}|+{\rm H.c.}
\end{equation}
with the phase difference $\phi_{jk}=\varphi_j-\varphi_k$ (assumed to be vanishing for simplicity) and the Rabi frequency
\begin{equation}\label{EffPara}
\Omega(t)=\frac{\Omega_1^2(t)}{\Delta_0+\Delta(t)/2}\approx\frac{\Omega_1^2(t)}{\Delta_0},
\end{equation}
where the approximation holds under $\Delta_0\gg\Delta(t)$. In the derivation of Eq.~(\ref{HameffMZ}), the first exponential terms in Eq.~(\ref{HamMrot}) have been neglected under the rotating wave approximation. As for the products by the second exponential terms, only those proportional to $\exp[i\int_0^t\Delta(s)ds]$ are slowly oscillating with time and then can be retained. In the double-rotated picture with respect to $H_{\rm MZM}=\Delta(t)(|\gamma_{1,k}\rangle\langle\gamma_{1,k}|-|\gamma_{1,j}\rangle\langle\gamma_{1,j}|)/2$, we have
\begin{equation}\label{HameffMZRot}
\begin{aligned}
H_{\rm eff}'(t)&=\frac{\Delta(t)}{2}\left(|\gamma_{1,j}\rangle\langle\gamma_{1,j}|
-|\gamma_{1,k}\rangle\langle\gamma_{1,k}|\right)\\
&+\Omega(t)|\gamma_{1,j}\rangle\langle\gamma_{1,k}|+{\rm H.c.},
\end{aligned}
\end{equation}
whose level transition diagram is shown in Fig.~\ref{model}(b). $\Delta(t)$ in Eq.~(\ref{HameffMZRot}) is directly controlled by the time-dependent detunings in laboratory, i.e., $\Delta_j(t)$ and $\Delta_k(t)$, rather than the extra driving fields in the previous protocol for effective Hamiltonian~\cite{Jin2025Rydberg}.

Using the universal quantum control theory~\cite{Jin2025Universal,Jin2025Entangling,Jin2025ErrCorr,Jin2025Rydberg} for an effective two-level system, the dynamics of the two MZMs $\gamma_{1,j}$ and $\gamma_{1,k}$ can be described in the ancillary representation spanned by
\begin{equation}\label{anci}
\begin{aligned}
|\mu_1(\theta,\alpha)\rangle&=\cos\theta(t)e^{i\frac{\alpha(t)}{2}}|\gamma_{1,j}\rangle
-\sin\theta(t)e^{-i\frac{\alpha(t)}{2}}|\gamma_{1,k}\rangle,\\
|\mu_2(\theta,\alpha)\rangle&=\sin\theta(t)e^{i\frac{\alpha(t)}{2}}|\gamma_{1,j}\rangle
+\cos\theta(t)e^{-i\frac{\alpha(t)}{2}}|\gamma_{1,k}\rangle,\\
\end{aligned}
\end{equation}
where the parameters $\theta(t)$ and $\alpha(t)$ manipulate the populations and the relative phase of the states, respectively. Substituting the ancillary basis states in Eq.~(\ref{anci}) to the von Neumann equation (\ref{von}) with the effective Hamiltonian (\ref{HameffMZRot}), we have
\begin{equation}\label{Condition}
\begin{aligned}
&\Delta(t)=\dot{\alpha}(t)+2\Omega(t)\cot2\theta(t)\cos\alpha(t),\\
&\Omega(t)=-\frac{\dot{\theta}(t)}{\sin\alpha(t)}.
\end{aligned}
\end{equation}
Using Eq.~(\ref{Condition}), the ancillary basis states in Eq.~(\ref{anci}) can be activated as useful nonadiabatic passages. Then due to Eq.~(\ref{U0}), the effective evolution operator can be parametrized as
\begin{equation}\label{U}
\begin{aligned}
U(\theta,\alpha)&=e^{if(\theta,\alpha)}|\mu_1(\theta,\alpha)\rangle\langle\mu_1(\theta_0,\alpha_0)|\\
&+e^{-if(\theta,\alpha)}|\mu_2(\theta,\alpha)\rangle\langle\mu_2(\theta_0,\alpha_0)|.
\end{aligned}
\end{equation}
Here the subscript $0$ indicates the initial boundary condition and the global phase $f(\theta,\alpha)$ is determined by
\begin{equation}\label{phase}
\dot{f}(\theta,\alpha)=\Omega(t)\frac{\cos\alpha(t)}{\sin2\theta(t)}=-\dot{\theta}(t)\frac{\cot\alpha(t)}{\sin2\theta(t)}.
\end{equation}
The second equivalence in the last equation used the second condition in Eq.~(\ref{Condition}). To avoid the singularity of the parameters $\Delta(t)$ and $\Omega(t)$ in Eq.~(\ref{Condition}), one can regard $\theta(t)$, $\alpha(t)$, and $f(\theta,\alpha)$ in Eq.~(\ref{phase}) as independent variables. The conditions in Eq.~(\ref{Condition}) can then be expressed as
\begin{subequations}\label{ConditionInv}
\begin{align}
&\Delta(t)=\dot{\alpha}(t)+2\dot{f}(\theta,\alpha)\cos2\theta(t), \label{Deltar} \\
&\dot{\alpha}(t)=-\frac{\ddot{\theta}\dot{f}\sin2\theta-\ddot{f}\dot{\theta}\sin2\theta
-2\ddot{f}\dot{\theta}^2\cos2\theta}{\dot{f}^2\sin^22\theta+\dot{\theta}^2},\label{dotalpha}\\
&\Omega(t)=-\sqrt{\dot{\theta}(t)^2+\dot{f}(\theta,\alpha)^2\sin^22\theta(t)}.\label{Om0}
\end{align}
\end{subequations}
With Eq.~(\ref{dotalpha}), $f(\theta,\alpha)$ can be simplified as $f(\theta)$.

\begin{figure}[htbp]
\centering
\includegraphics[width=0.9\linewidth]{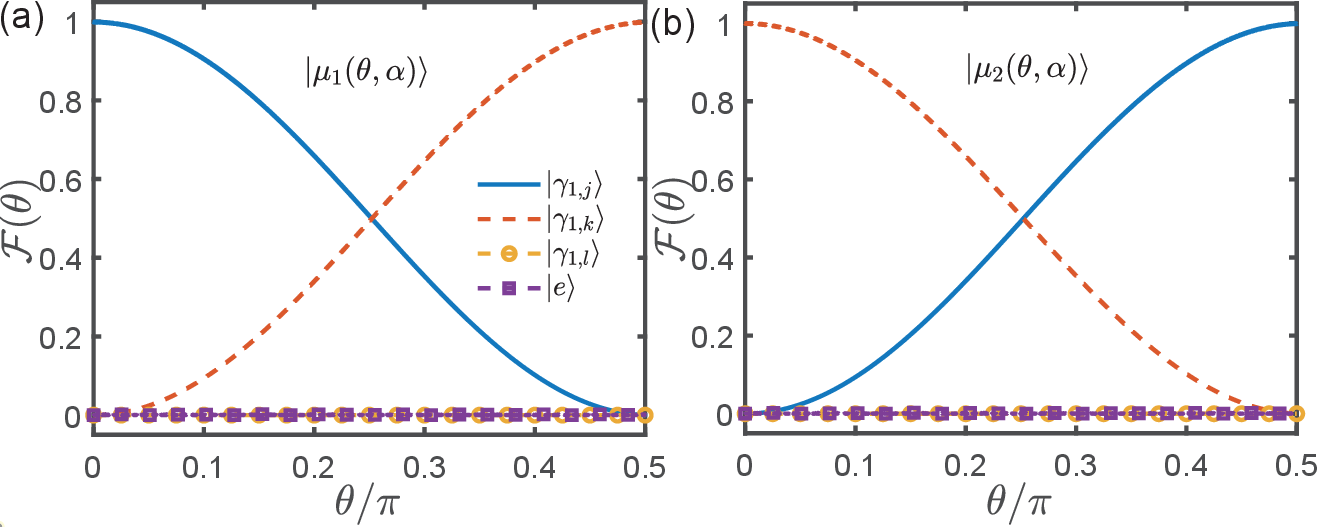}
\caption{Fidelity $\mathcal{F}(\theta)$ for the states $|n\rangle$, $n=\gamma_{1,j},\gamma_{1,k},\gamma_{1,l},e$, as a function of the parameter $\theta(t)$ during the braiding process with the initial population in MZMs (a) $\gamma_{1,j}$ and (b) $\gamma_{1,k}$. $\Delta_j(t)=\Delta_0-\Delta(t)/2$ and $\Delta_k(t)=\Delta_0+\Delta(t)/2$ where $\Delta(t)$ in Eq.~(\ref{Deltar}) and it is in the order of $\Delta(t)\sim10^{-2}\Delta_0$. $\Delta_l=2\Delta_0$. Under $\Omega_1(t)=\Omega_2(t)=\Omega_3(t)$, the effective Rabi frequency $\Omega(t)$ satisfies Eqs.~(\ref{EffPara}) and (\ref{Om0}) with $\theta(t)=\pi t/(2T)$ ($T$ is the control period) and $f(\theta,\alpha)=0$. The Rabi frequency is in the order of $\Omega_1(t)\sim 10^{-2}\Delta_0$.}\label{Braidp}
\end{figure}

Equations~(\ref{anci}) and (\ref{U}) indicate that the system can evolve along either $|\mu_1(\theta,\alpha)\rangle$ or $|\mu_2(\theta,\alpha)\rangle$, depending on the boundary conditions of the parameters $\theta(t)$, $\alpha(t)$, and $f(\theta,\alpha)$. For example, when $\theta(0)=0$, $\theta(T)=\pi/2$ with $T$ the evolution period, and $\alpha(t)=\pi$, the evolution operator in Eq.~(\ref{U}) can become
\begin{equation}\label{UB0}
U(\theta,\alpha)=-e^{if(\theta,\alpha)}|\gamma_{1,k}\rangle\langle\gamma_{1,j}|
+e^{-if(\theta,\alpha)}|\gamma_{1,j}\rangle\langle\gamma_{1,k}|.
\end{equation}
When $f(\theta,\alpha)=0$, it is further reduced to a typical braiding operation between the two MZMs $\gamma_{1,j}$ and $\gamma_{1,k}$~\cite{Ivanov2001NonAbelian,Stone2006Fusion,Read2009NonAbelian,Sau2011Controlling,
vanHeck2012Coulomb,Kraus2013Braiding}:
\begin{equation}\label{UB}
U(\theta,\alpha)=-|\gamma_{1,k}\rangle\langle\gamma_{1,j}|+|\gamma_{1,j}\rangle\langle\gamma_{1,k}|
=\exp\left(\frac{\pi}{4}\gamma_{1,j}\gamma_{1,k}\right).
\end{equation}

Due to the transition diagram in the inset of Fig.~\ref{model}(a), the performance of our braiding protocol can be evaluated by the fidelities for all the four relevant states: $\mathcal{F}(\theta)\equiv|\langle n|\psi(\theta)\rangle|^2$, $n=\gamma_{1,j},\gamma_{1,k},\gamma_{1,l},e$, where the wave function $|\psi(\theta)\rangle$ is directly calculated by the Schr\"odinger equation $i\partial_t|\psi(\theta)\rangle=H_{\rm tot}(t)|\psi(\theta)\rangle$ with the total Hamiltonian~(\ref{HamDri}). Figures~\ref{Braidp}(a) and (b) present the numerical results of the fidelities starting from different MZMs. They are exactly the same as the braiding process controlled by $U(\theta,\alpha)$ in Eq.~(\ref{UB}), during which the states $|\gamma_{1,l}\rangle$ and $|e\rangle$ are effectively decoupled from the whole evolution. Particularly, in Fig.~\ref{Braidp}(a), the initial state $|\gamma_{1,j}\rangle$ evolves into an equal superposition of the states $|\gamma_{1,j}\rangle$ and $|\gamma_{1,k}\rangle$ when $\theta=\pi/4$ and then completely transfers to the state $|\gamma_{1,k}\rangle$ when $\theta=\pi/2$. When the system is initially populated on MZM $\gamma_{1,k}$, it can be faithfully transferred to the target MZM $\gamma_{1,j}$ when $\theta(T)=\pi/2$ as shown in Fig.~\ref{Braidp}(b). In the languages of UQC, Figures~\ref{Braidp}(a) and (b) simulate the nonadiabatic and transitionless passages $|\mu_1\rangle$ and $|\mu_2\rangle$ in Eq.~(\ref{anci}), respectively.

Our protocol can be straightforwardly generalized from $3$ MZMs to $M$ MZMs. In this case, the control Hamiltonian~(\ref{HamMrot}) will be rescaled to
\begin{equation}\label{HamMrotN}
\begin{aligned}
H_I^{(M)}(t)&=\sum_{k=1}^M\Omega_k(t)\Big\{e^{i\int_0^t[\omega_k(s)+\omega]ds+i\varphi_k}\\
&+e^{-i\int_0^t\Delta_k(s)ds-i\varphi_k}\Big\}|e\rangle\langle\gamma_{1,k}|+{\rm H.c.}.
\end{aligned}
\end{equation}
To modulate the hopping and pairing interaction between LD and the end sites of $M$ Kitaev chains, a roughly linear-scaling experimental overhead has to be paid to control the eigenfrequencies of the fermions and their neighboring couplings in the Kitaev chains, including $\mu$, $J$, and $g$. Then the preceding construction of braiding operation can be implemented between any selected pair of MZMs, e.g., the $j$th and $k$th MZMs, $j\neq k$. It is found that under the isotropic condition for Rabi frequencies $\Omega_k(t)=\Omega_1(t)$ with $1\leq k\leq M$, the large detuning condition $\Delta_j(t),\Delta_k(t),|\Delta_j(t)-\Delta_{l\neq j,k}(t)|,|\Delta_k(t)-\Delta_{l\neq j,k}(t)|\gg\Omega_1(t)$, and the near-resonant condition $|\Delta_j(t)-\Delta_k(t)|\sim\Omega_1(t)$, one can again obtain the same two-body effective Hamiltonian as in Eq.~(\ref{HameffMZ}) by Eq.~(\ref{Heff}).

\subsection{Suppress systematic errors in local defect}\label{Correct}

While the braiding operations of MZMs are resilient to the local perturbations in the other Majorana fermions, this topological protection does not cover the systematic errors about the mediator LD. As indicated by Eq.~(\ref{HamrotNon}), the unavoidable systematic errors associated with LD can cause undesirable transitions among the universal passages. A reliable mitigating method is incorporating an error correction mechanism through e.g., Eq.~(\ref{OptGeneralOne}) to our protocol during the braiding operations of MZMs.

From a pedagogical perspective, the full Hamiltonian $H_{\rm rot}(t)$ in Eq.~(\ref{HamDri}) can be perturbed by: (i) the fluctuation in the splitting frequency of the local defect, i.e.,
\begin{equation}\label{HamDriLocal}
\omega\rightarrow(1+\epsilon)\omega,
\end{equation}
or (ii) the Rabi-frequency fluctuation in the driving field for one of the transitions $|e\rangle\leftrightarrow|\gamma_{1,1}\rangle$, i.e.,
\begin{equation}\label{HamDriGlob}
\Omega_1(t)\rightarrow(1+\epsilon)\Omega_1(t),
\end{equation}
where $\epsilon$ is the perturbative coefficient measuring the error magnitude. To suppress the undesirable transitions between the passages by our correction mechanism~(\ref{OptGeneralOne}), one can simply set the global phase as
\begin{equation}\label{phasef}
\frac{\partial f(\theta)}{\partial\theta(t)}=\lambda,
\end{equation}
where $\lambda$ scales the variation rate of $f(\theta)$ with respect to $\theta(t)$. $\lambda=0$ represents the parallel-transport condition with a static global phase~\cite{Jin2025ErrCorr} and $|\lambda|\gg1$ describes a rapid-varying global-phase that can suppress the systematic errors in the relevant nonadiabatic passage. Indicated by Eqs.~(\ref{Condition}) and (\ref{phase}), the global phase is uniquely determined by $\Delta(t)$ and $\Omega(t)$, through the parameters $\theta(t)$ and $\alpha(t)$ in universal passage. Experimentally, the rapid modulation of the global phase can be achieved by the gate voltages of the superconductor circuit embedded in the hybrid semiconductor-superconductor systems~\cite{Dvir2023Realization,Shi2024Probing,Bordin2025Enhanced,Microsoft2025Interferometric,
Nitsch2025Poor}, where the superconductor circuit is adiabatically eliminated to effectively realize the tunable hopping and pairing couplings between QDs as described by Eq.~(\ref{HamDriRedu}).

\begin{figure}[htbp]
\centering
\includegraphics[width=0.9\linewidth]{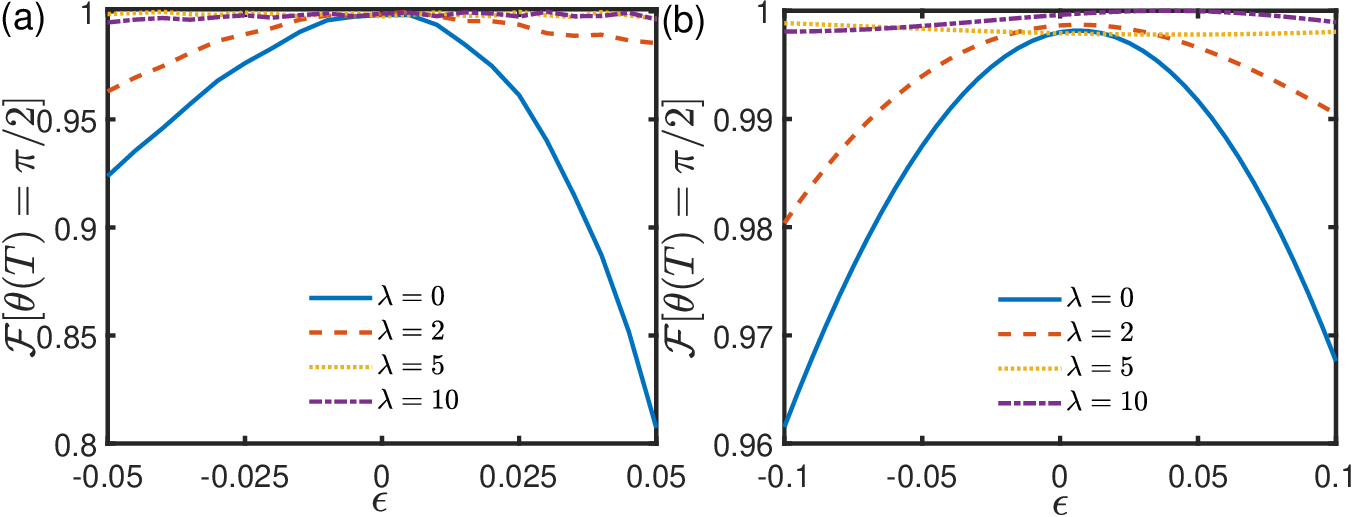}
\caption{Fidelity $\mathcal{F}[\theta(T)=\pi/2]$ about the target state $|\gamma_{1,k}\rangle$ versus the perturbative coefficient $\epsilon$, controlled by the full Hamiltonian $H_{\rm tot}(t)$~(\ref{HamDri}) with (a) deviated transition frequency $\omega$ in Eq.~(\ref{HamDriLocal}) and (b) deviated Rabi frequency $\Omega_1(t)$ in Eq.~(\ref{HamDriGlob}), under the global phase $f(\theta)$ in Eq.~(\ref{phasef}) with various coefficients $\lambda$. The other parameters are the same as Fig.~\ref{Braidp}(a).}\label{ComErr}
\end{figure}

Figures~\ref{ComErr}(a) and (b) demonstrate the fidelity $\mathcal{F}[\theta(T)=\pi/2]$ about the target state $|\gamma_{1,k}\rangle$ versus the dimensionless error magnitude $\epsilon$, in the presence of the systematic errors in Eqs.~(\ref{HamDriLocal}) and (\ref{HamDriGlob}), respectively. One can find that the braiding operations incorporating with our correction mechanism ($|\lambda|>1$) exhibits superior performances in comparison to those with no correction mechanism ($\lambda=0$). In Fig.~\ref{ComErr}(a) for $\lambda=0$, the fidelity declines to $\mathcal{F}=0.924$ when $\epsilon=-0.05$ and $\mathcal{F}=0.807$ when $\epsilon=0.05$. Under our correction mechanism, the fidelity admits a lower bound as $\mathcal{F}>0.963$ when $\lambda=2$ and $\mathcal{F}>0.995$ when $\lambda\geq5$, in the whole range of $\epsilon\in[-0.05,0.05]$. As shown in Fig.~\ref{ComErr}(b), the braiding operation with our correction mechanism exhibits a more significant robustness against the deviation in Rabi frequency of a local driving field. In particular, when $\lambda=0$, the fidelity is about $\mathcal{F}=0.962$ when $\epsilon=-0.1$ and $\mathcal{F}=0.968$ when $\epsilon=0.1$. The adverse effects from the local errors can be suppressed so that $\mathcal{F}>0.980$ when $\lambda=2$, and $\mathcal{F}>0.998$ when $\lambda\geq5$.

\subsection{Suppress quasiparticle poisoning}\label{quasipoison}

The fermion parity of the subspace of Majorana zero modes is conserved with the systematic errors on local defect as discussed in Sec.~\ref{Correct}. However, in practical topological quantum computation, the fermion parity can be broken by quasiparticle poisoning~\cite{Flsensberg2010Tunneling,Leijnse2011Scheme,Rainis2012Majorana,Karzig2021Quasiparticle}. It occurs when an undesired Majorana fermion participates the braiding operation and typically arises from the nonideal conditions of the system, including environmental noise, perturbations in system parameters, and thermal excitations~\cite{Karzig2021Quasiparticle}.

We consider the following Hamiltonian to simulate the effect from the quasiparticle poisoning~\cite{Knapp2016Nature}:
\begin{equation}\label{Hampoison}
\tilde{H}_{\rm tot}(t)=H_{\rm eff}(t)+H_{\rm err}(t),
\end{equation}
where $H_{\rm eff}(t)$ is given by Eq.~(\ref{HameffMZ}) with $\Omega(t)$ the effective Rabi-frequency of the transition between the interested states $|\gamma_{1,j}\rangle$ and $|\gamma_{1,k}\rangle$, and the error Hamiltonian $H_{\rm err}(t)$ is given by
\begin{equation}\label{Herrpoison}
H_{\rm err}(t)=\epsilon\Omega(t)|\gamma_{1,j}\rangle\langle\gamma_{1,l}|+{\rm H.c.},
\end{equation}
with $\epsilon$ the leakage magnitude out of the topological subspace spanned by $|\gamma_{1,j}\rangle$ and $|\gamma_{1,k}\rangle$. The quasiparticle poisoning described by Eq.~(\ref{Herrpoison}) can arise from the unwanted hopping and pairing interactions between the fermions at the ends of the $j$th and $l$th Kitaev chains, i.e., $c_{1,j}$ and $c_{1,l}$.

\begin{figure}[htbp]
\centering
\includegraphics[width=0.8\linewidth]{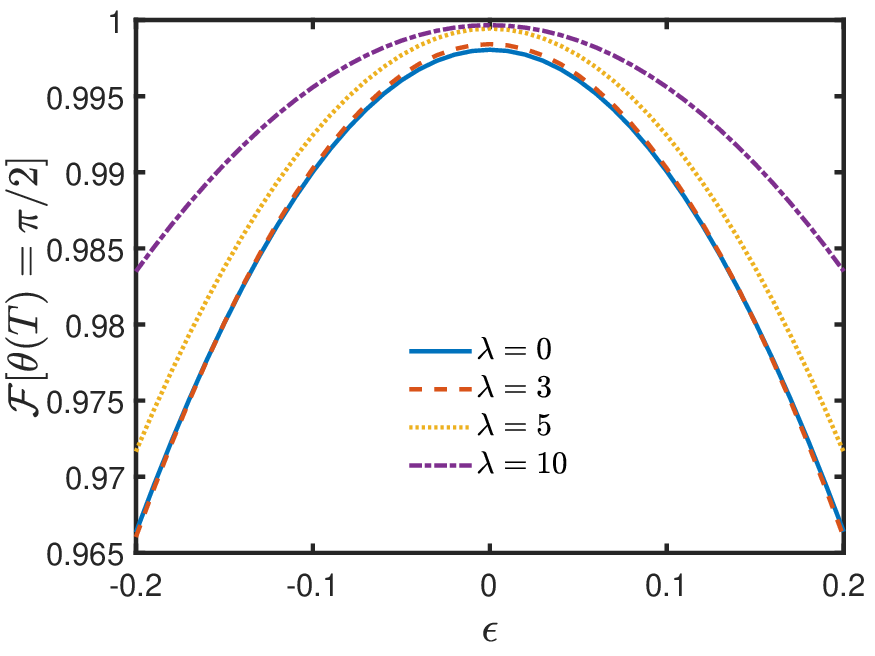}
\caption{Fidelity $\mathcal{F}[\theta(T)=\pi/2]$ about the target state $|\gamma_{1,k}\rangle$ versus the perturbative coefficient $\epsilon$, controlled by the nonideal Hamiltonian $\tilde{H}_{\rm tot}(t)$ in Eq.~(\ref{Hampoison}). The other parameters are the same as Fig.~\ref{Braidp}(a).}\label{poison}
\end{figure}

In Fig.~\ref{poison}, we demonstrate the fidelity $\mathcal{F}[\theta(T)=\pi/2]$ about the target state $|\gamma_{1,k}\rangle$ versus the dimensionless error magnitude $\epsilon$, under the nonideal Hamiltonian~(\ref{Hampoison}). One can find that our correction mechanism in Eq.~(\ref{phasef}) is still applicable to suppress the effect from the quasiparticle poisoning. In particular, for $\lambda=0$, the fidelity declines to $\mathcal{F}[\theta(T)=\pi/2]=0.967$ when $\epsilon=\pm0.2$. When $\lambda=5$, the fidelity is enhanced to $\mathcal{F}[\theta(T)=\pi/2]\geq0.973$ in the whole range of $\epsilon\in[-0.2,0.2]$; and when $\lambda=10$, the lower-bound of the fidelity is over $0.984$ within the range of $\epsilon\in[-0.2,0.2]$.

Moreover, our protocol can be alternatively combined with the measurement-based error correction~\cite{Knapp2016Nature} to mitigate the quasiparticle poisoning. Particularly, the projection operator~\cite{Knapp2016Nature} for the $j$th and $k$th MZMs in our model can be defined as
\begin{equation}\label{project}
\Pi_\pm=\frac{1\pm i\gamma_{1,j}\gamma_{1,k}}{2}.
\end{equation}
In the presence of the leakage out of the topological subspace due to the unwanted crosstalk similar to Eq.~(\ref{Herrpoison}), the ideal braid operation $U(\theta,\alpha)$ in Eq.~(\ref{UB}) becomes
\begin{equation}\label{Uepsi}
\begin{aligned}
&U(\theta,\alpha)\rightarrow U_\epsilon=e^{\frac{\pi}{4}\gamma_{1,j}\gamma_{1,k}+\epsilon\gamma_{1,j}\gamma_{1,l}}\\
&\approx e^{\frac{\pi}{4}\gamma_{1,j}\gamma_{1,k}}e^{\epsilon\gamma_{1,j}\gamma_{1,l}}e^{-\frac{1}{2}
[\frac{\pi}{4}\gamma_{1,j}\gamma_{1,k}, \epsilon\gamma_{1,j}\gamma_{1,l}]},\\
&\approx U(\theta,\alpha)\left(1+\epsilon\gamma_{1,j}\gamma_{1,l}-\frac{\pi\epsilon}{8}
[\gamma_{1,j}\gamma_{1,k}, \gamma_{1,j}\gamma_{1,l}]\right)+\mathcal{O}(\epsilon^2),\\
&=U(\theta,\alpha)\left(1+\epsilon\gamma_{1,j}\gamma_{1,l}-\frac{\pi\epsilon}{4}\gamma_{1,k}\gamma_{1,l}\right)
+\mathcal{O}(\epsilon^2).
\end{aligned}
\end{equation}
Here we have applied the Baker-Campbell-Hausdorff formula with a small perturbation $\epsilon\ll1$. Assume that the initial state of the system $|\psi_i\rangle$ satisfies the parity condition $i\gamma_{1,j}\gamma_{1,k}|\psi_i\rangle=1$. Then under the combination strategy, the final state justifies that the measurement-based error correction $\Pi_+$ in Eq.~(\ref{project}) can be used to suppress the leakage out of the topological subspace spanned by $\gamma_{1,j}$ and $\gamma_{1,k}$ to the first order:
\begin{equation}\label{final}
\begin{aligned}
&|\psi_f\rangle=\Pi_+U_\epsilon|\psi_i\rangle=\Pi_+U(\theta,\alpha)\Big(1+\epsilon\gamma_{1,j}\gamma_{1,l}\\
&-\frac{\pi\epsilon}{4}\gamma_{1,k}\gamma_{1,l}\Big)|\psi_i\rangle+\mathcal{O}(\epsilon^2)
=U(\theta,\alpha)|\psi_i\rangle+\epsilon\Big[U(\theta,\alpha)\\
&\times\Pi_+\gamma_{1,j}\gamma_{1,l}-\frac{\pi}{4}U(\theta,\alpha)\Pi_+\gamma_{1,k}\gamma_{1,l}\Big]|\psi_i\rangle
+\mathcal{O}(\epsilon^2)\\ &=U(\theta,\alpha)|\psi_i\rangle
+\epsilon\Big[U(\theta,\alpha)\gamma_{1,j}\gamma_{1,l}\Pi_--\frac{\pi}{4}U(\theta,\alpha)\\
&\times\gamma_{1,k}\gamma_{1,l}\Pi_-\Big]|\psi_i\rangle+\mathcal{O}(\epsilon^2)
=U(\theta,\alpha)|\psi_i\rangle+\mathcal{O}(\epsilon^2).
\end{aligned}
\end{equation}
In this derivation, we have used the anti-commutation relations $\{\gamma_{1,j}\gamma_{1,k}, \gamma_{1,j}\gamma_{1,l}\}=0$ and $\{\gamma_{1,j}\gamma_{1,k}, \gamma_{1,k}\gamma_{1,l}\}=0$, and the result $\Pi_-|\psi_i\rangle=0$.

\section{Chiral population transfer among Majorana zero modes}\label{chiral}

\begin{figure}[htbp]
\centering
\includegraphics[width=0.9\linewidth]{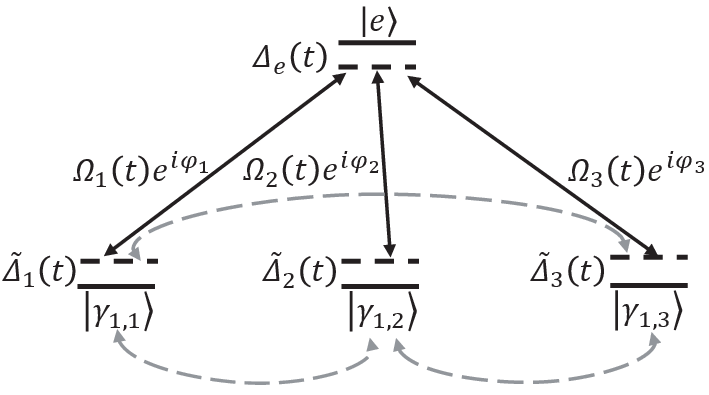}
\caption{Four-level transition diagram with the coupling between LD and MZMs $\gamma_{1,k}$, $k=1,2,3$. Black solid lines describe the off-resonant driving fields on the transition $|e\rangle\leftrightarrow|\gamma_{1,k}\rangle$ with the detuning $\Delta_e(t)-\tilde{\Delta}_k(t)$, the Rabi-frequency $\Omega_k(t)$, and the phase $\varphi_k$. The gray dashed lines imply the population transfer among MZMs along a clockwise or counterclockwise direction.}\label{modelchiral}
\end{figure}

Beyond the large-detuning regime, we can apply the universal quantum control theory~\cite{Jin2025Universal} (see Appendix~\ref{UniPara} for details) to realize a chiral population transfer~\cite{Wang2016Mesoscopic,Liu2020Synthesizing,Qi2022Chiral} among the three MZMs $\gamma_{1,1}$, $\gamma_{1,2}$, and $\gamma_{1,3}$ without elimination of LD. As shown in Fig.~\ref{modelchiral}, a LD coupled to MZM $\gamma_{1,k}$'s, $k=1,2,3$, by the local driving field with the detuning $\Delta_e(t)-\tilde{\Delta}_k(t)$, the Rabi-frequency $\Omega_k(t)$, and the phase $\varphi_k$. The full Hamiltonian can be obtained by Eq.~(\ref{HamMrot}), which is recalled as below
\begin{equation}\label{HamMrotR}
\begin{aligned}
H_I(t)&=\sum_{k=1}^3\Omega_k(t)\Big\{e^{i\int_0^t[\omega_k(s)+\omega]ds+i\varphi_k}\\
&+e^{-i\int_0^t[\omega_k(s)-\omega]ds-i\varphi_k}\Big\}|e\rangle\langle\gamma_{1,k}|+{\rm H.c.}.
\end{aligned}
\end{equation}
Formally the driving frequencies satisfy $\omega_k(t)=-\omega+\Delta_e(t)-\tilde{\Delta}_k(t)$. In the large transition-frequency regime $\omega\gg\Omega_k(t),\Delta_e(t),\tilde{\Delta}_k(t)$, $1\le k\le3$, Eq.~(\ref{HamMrotR}) can be reduced to
\begin{equation}\label{HamMrotfour}
H_I(t)=\sum_{k=1}^3\Omega_k(t)e^{i\int_0^t[-\tilde{\Delta}_k(s)+\Delta_e(s)]ds+i\varphi_k}|e\rangle\langle\gamma_{1,k}|+{\rm H.c.}
\end{equation}
under the rotating-wave approximation. In the double-rotated frame with respect to $H_0=-\Delta_e(t)|e\rangle\langle e|-\sum_{k=1}^3\tilde{\Delta}_k(t)|\gamma_{1,k}\rangle\langle\gamma_{1,k}|$, the Hamiltonian (\ref{HamMrotfour}) can be transformed as
\begin{equation}\label{Hamfour}
\begin{aligned}
H_I(t)&=\Delta_e(t)|e\rangle\langle e|+\sum_{k=1}^{3}\Big\{\tilde{\Delta}_k(t)|\gamma_{1,k}\rangle\langle\gamma_{1,k}|\\
&+\left[\Omega_k(t)e^{i\varphi_k}|e\rangle\langle\gamma_{1,k}|+{\rm H.c.}\right]\Big\}.
\end{aligned}
\end{equation}

Using the UQC recipe for two-band systems~\cite{Jin2025Entangling}, the system dynamics can be described in the ancillary picture by four ancillary basis states, which can be chosen as
\begin{equation}\label{AnciFour}
\begin{aligned}
|\mu_1(t)\rangle&=\cos\theta(t)|\gamma_{1,1}\rangle-\sin\theta(t)e^{-i\alpha_1(t)}|\gamma_{1,2}\rangle,\\
|\mu_2(t)\rangle&=\cos\phi(t)|b_1(t)\rangle-\sin\phi(t)e^{-i\alpha_2(t)}|\gamma_{1,3}\rangle,\\
|\mu_3(t)\rangle&=\cos\chi(t)|b_2(t)\rangle-\sin\chi(t)e^{-i\alpha_3(t)}|e\rangle,\\
|\mu_4(t)\rangle&=\sin\chi(t)|b_2(t)\rangle+\cos\chi(t)e^{-i\alpha_3(t)}|e\rangle.
\end{aligned}
\end{equation}
Here the bright states are given by
\begin{equation}\label{brightFour}
\begin{aligned}
|b_1(t)\rangle&=\sin\theta(t)|\gamma_{1,1}\rangle+\cos\theta(t)e^{-i\alpha_1(t)}|\gamma_{1,2}\rangle,\\
|b_2(t)\rangle&=\sin\phi(t)|b_1(t)\rangle+\cos\phi(t)e^{-i\alpha_2(t)}|\gamma_{1,3}\rangle,
\end{aligned}
\end{equation}
which hold the orthogonality with the relevant ancillary basis states, i.e., $\langle\mu_n(t)|b_n(t)\rangle=0$, $n=1,2$. The time-dependent parameters $\theta(t)$, $\phi(t)$, and $\chi(t)$ control the state populations, whereas $\alpha_1(t)$, $\alpha_2(t)$, and $\alpha_3(t)$ determine the relative phases among states. Equation~(\ref{AnciFour}) constitutes a completed and orthornormal basis for the Hilbert space spanned by LD excited state and three MZMs.

Although each basis state in Eq.~(\ref{AnciFour}) serves as a nonadiabatic passage when satisfying the von Neumann equation~(\ref{von}) with the Hamiltonian $H(t)$~(\ref{Hamfour}), it is found that only the last two can realize a perfect chiral transfer of MZMs. Without loss of generality, we substitute $|\mu_4(t)\rangle$ in Eq.~(\ref{AnciFour}) to the von Neumann equation~(\ref{von}) and then obtain the conditions for the Rabi-frequencies
\begin{equation}
\begin{aligned}\label{CondRabi}
\Omega_1(t)&=-\Big(\dot{\chi}\sin\phi\sin\theta+\dot{\phi}\tan\chi\cos\phi\sin\theta+\dot{\theta}\tan\chi\\
&\times\sin\phi\cos\theta\Big)/\sin(\varphi_1+\alpha_3),\\
\Omega_2(t)&=-\Big(\dot{\chi}\sin\phi\cos\theta+\dot{\phi}\tan\chi\cos\phi\cos\theta-\dot{\theta}\tan\chi\\
&\times\sin\phi\sin\theta\Big)/\sin(\varphi_2-\alpha_1+\alpha_3),\\
\Omega_3(t)&=-\left(\dot{\chi}\cos\phi-\dot{\phi}\tan\chi\sin\phi\right)/\sin(\varphi_3-\alpha_2+\alpha_3),\\
\end{aligned}
\end{equation}
and those for the detunings
\begin{equation}
\begin{aligned}\label{CondDetu}
&\tilde{\Delta}_1(t)=-\Omega_1(t)\frac{\cot\chi}{\sin\phi\sin\theta}\cos(\varphi_1+\alpha_3),\\
&\tilde{\Delta}_2(t)=\dot{\alpha}_1-\Omega_2(t)\frac{\cot\chi}{\sin\phi\cos\theta}\cos(\varphi_2-\alpha_1+\alpha_3)\\
&\tilde{\Delta}_3(t)=\dot{\alpha}_2-\Omega_3\frac{\cot\chi}{\cos\phi}\cos(\varphi_3-\alpha_2+\alpha_3),\\
&\Delta_e(t)=\dot{\alpha}_3-\Omega_1(t)\tan\chi\sin\phi\sin\theta\cos(\varphi_1+\alpha_3)\\
&-\Omega_2(t)\tan\chi\sin\phi\cos\theta\cos(\varphi_2-\alpha_1+\alpha_3)\\ &-\Omega_3\tan\chi\cos\phi\cos(\varphi_3-\alpha_2+\alpha_3).
\end{aligned}
\end{equation}
With Eqs.~(\ref{CondRabi}) and (\ref{CondDetu}), the system evolution can be precisely manipulated to follow the desired passage. A complete loop of chiral population transfer of the three MZMs can be divided into three sequential stages, each lasing an identical time interval $T$. In particular, for the system initialized in the state $|\gamma_{1,1}\rangle$, the clockwise transfer proceeds through stage (i) $|\gamma_{1,1}\rangle\rightarrow|\gamma_{1,2}\rangle$ when $t\in[0, T]$, stage (ii) $|\gamma_{1,2}\rangle\rightarrow|\gamma_{1,3}\rangle$ when $t\in[T, 2T]$, and stage (iii) $|\gamma_{1,3}\rangle\rightarrow|\gamma_{1,1}\rangle$ when $t\in[2T, 3T]$. They are determined by the boundary conditions for the parameters $\theta(t)$, $\phi(t)$, and $\chi(t)$: $\theta(0)=\phi(0)=\chi(0)=\pi/2$, $\theta(T)=\pi$, $\phi(T)=\chi(T)=\pi/2$, $\phi(2T)=0$, $\chi(2T)=\pi/2$, and $\theta(3T)=\phi(3T)=\chi(3T)=\pi/2$. For our purpose, $\theta(t)$ and $\phi(t)$ for stages (i), (ii), and (iii) of the $k$th loop, $k\geq1$, can be set as
\begin{equation}\label{CondTPC}
\begin{aligned}
\theta(t)&=\Phi(t)+\frac{\pi}{2},\quad \phi(t)=2\Phi(t)+\frac{\pi}{2},\\
\theta(t)&=\Phi(t)+\frac{\pi}{2},\quad \phi(t)=\Phi(t),\\
\theta(t)&=\phi(t)=\Phi(t),
\end{aligned}
\end{equation}
respectively, with $\Phi(t)\equiv\pi[t-3(k-1)T]/(2T)$, and $\chi(t)$ for the whole loop can be set as
\begin{equation}\label{CondChi}
\chi(t)=\frac{\pi}{2}\left[1+\frac{\sin{(\frac{\pi t}{T})}}{1+\sin^2(\frac{\pi t}{T})}\right].
\end{equation}

\begin{figure}[htbp]
\centering
\includegraphics[width=0.9\linewidth]{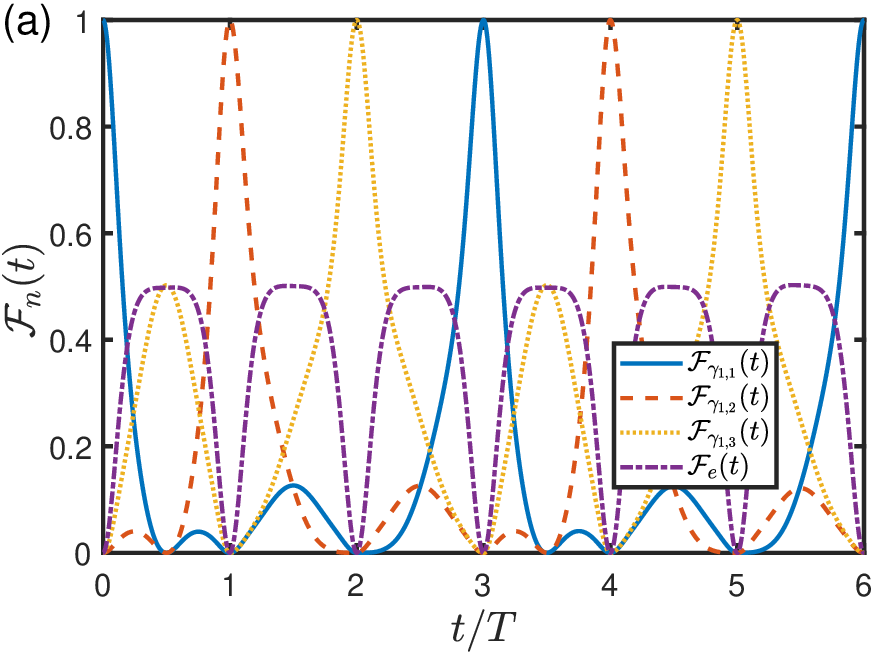}
\includegraphics[width=0.9\linewidth]{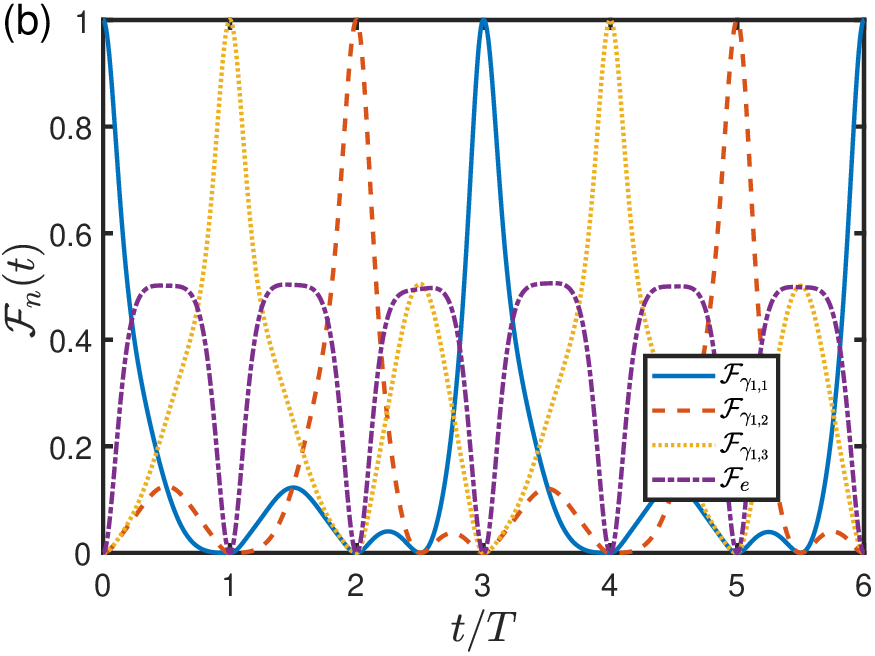}
\caption{Fidelity dynamics $\mathcal{F}_n(t)$ about various states $|n\rangle$, $n=\gamma_{1,1},\gamma_{1,2},\gamma_{1,3},e$, for two loops of (a) clockwise and (b) counterclockwise population transfers. The Rabi frequencies and the detunings of the Hamiltonian~(\ref{Hamfour}) are set by Eqs.~(\ref{CondRabi}) and (\ref{CondDetu}) with $\varphi_1+\alpha_3=\pi/2$, $\varphi_2-\alpha_1+\alpha_3=\pi/2$, $\varphi_3-\alpha_2+\alpha_3=\pi/2$ and (a) $\theta(t)$, $\phi(t)$, and $\chi(t)$ in Eqs.~(\ref{CondTPC}) and (\ref{CondChi}) and (b) $\theta(t)$, $\phi(t)$, and $\chi(t)$ in Eqs.~(\ref{CondTcount}), (\ref{CondPhicount}), and (\ref{CondChi}). The Rabi frequencies are in the same order as those in Fig.~\ref{Braidp}.}\label{Chiraltrans}
\end{figure}

Similarly, the counterclockwise population transfer proceeds through stage (i) $|\gamma_{1,1}\rangle\rightarrow|\gamma_{1,3}\rangle$, stage (ii) $|\gamma_{1,3}\rangle\rightarrow|\gamma_{1,2}\rangle$, and stage (iii) $|\gamma_{1,2}\rangle\rightarrow|\gamma_{1,1}\rangle$. For the $k$th loop, $\theta(t)$ for stage (i) and stage (ii-iii) can be set as
\begin{equation}\label{CondTcount}
\theta(t)=\Phi(t)+\frac{\pi}{2}, \quad \theta(t)=\Phi(t)+\pi,
\end{equation}
respectively, and $\phi(t)$ for stages (i-ii) and stage (iii) can be chosen as
\begin{equation}\label{CondPhicount}
\phi(t)=\Phi(t)+\frac{\pi}{2}, \quad \phi(t)=\Phi(t),
\end{equation}
respectively. Both $\chi(t)$ and $\Phi(t)$ hold the preceding definitions.

Figure~\ref{Chiraltrans} demonstrates the fidelity dynamics $\mathcal{F}_n(t)$ about the four relevant states $|n\rangle$, $n=\gamma_{1,1},\gamma_{1,2},\gamma_{1,3},e$, describing the chiral population transfer. It is found that the population can propagate with a unit fidelity at the ends of each stages in both clockwise and counterclockwise manners. In Fig.~\ref{Chiraltrans}(a) for the clockwise population transfer, the initial population on the first MZM $\gamma_{1,1}$ can be completely transferred to the second one $\gamma_{1,2}$ when $t=T$, despite the unwanted states are temporally occupied during stage (i). We have up to $\mathcal{F}_e\approx\mathcal{F}_{\gamma_{1,3}}\approx0.5$ when $t=0.5T$. At the end of stage (ii), the population transfers completely to the third MZM $\gamma_{1,3}$ at $t=2T$. And the loop is completed by the population returning to the state $|\gamma_{1,1}\rangle$ at $t=3T$. During the second loop $t\in[3T, 6T]$, we have another perfect chiral transfer in the clockwise manner. Figure~\ref{Chiraltrans}(b) demonstrates the counterclockwise population transfer with the sequences: the population in the state $|\gamma_{1,1}\rangle$ can be transferred to the state $|\gamma_{1,3}\rangle$ when $t=T$, followed by the state $|\gamma_{1,2}\rangle$ when $t=2T$, and then back to the state $|\gamma_{1,1}\rangle$ when $t=3T$. The second loop of the counterclockwise population transfer exhibits identical behavior of the first loop.

In conventional protocols for chiral transfer, both spin chirality~\cite{Liu2020Synthesizing} and bosonic-mode chirality~\cite{Qi2022Chiral} depend on an effective time-reversal-symmetry broken Hamiltonian, from either three-body interaction or Floquet engineering. Normally the Hamiltonian is obtained by the adiabatic elimination over ancillary state or mode in the large-detuning regime, resulting in an effective coupling strength (typically it is weak) among the interested subsystems. Thus, these protocols are susceptible to the environment noise due to the long running period. Moveover, they demand a precise control over the local phases of individual driving fields. In sharp contrast, our protocol enables the chiral population transfer in a nonadiabatic way. It does not invoke synthesizing an artificial magnetic field and does not require a particular phase $\varphi_n$ for each driving field.

Our protocol for chiral population transfer can also be scaled with a linear resource overhead in case of $M$ MZMs described by Eq.~(\ref{HamMrotN}). Particularly, by setting the driving frequencies as $\omega_k(t)=-\omega+\Delta_e(t)-\tilde{\Delta}_k(t)$ and the transition frequency $\omega\gg\Omega_k(t),\Delta_e(t),\tilde{\Delta}_k(t)$ for $1\leq k\leq M$, the Hamiltonian~(\ref{HamMrotN}) can become
\begin{equation}\label{HamfourN}
\begin{aligned}
H_I^{(M)}(t)&=\Delta_e(t)|e\rangle\langle e|+\sum_{k=1}^{M}\Big\{\tilde{\Delta}_k(t)|\gamma_{1,k}\rangle\langle\gamma_{1,k}|\\
&+\left[\Omega_k(t)e^{i\varphi_k}|e\rangle\langle\gamma_{1,k}|+{\rm H.c.}\right]\Big\}
\end{aligned}
\end{equation}
in the rotating frame with respect to $H_0=-\Delta_e(t)|e\rangle\langle e|-\sum_{k=1}^M\tilde{\Delta}_k(t)|\gamma_{1,k}\rangle\langle\gamma_{1,k}|$, under the rotating-wave approximation. For this $1+M$-dimensional model, the recipe in Ref.~\cite{Jin2025Entangling} has provided two ancillary basis states, as the superpositions of the LD excited state and MZMs, either of which can be activated as a useful passage for chiral population transfer.

\section{Conclusion and comparison}\label{conclusion}

In the theoretical framework of universal quantum control, we construct high-fidelity, nonadiabatic, fault-tolerant, and linear-scalable braiding operations for Majorana zero modes that are coupled to a common two-level local defect. The braiding operation between an arbitrary pair of MZMs is generated on the induced exchange interaction by using the largely detuned driving fields on LD and the subsequent rotating wave approximation. Our universal passage can employ a rapidly-varying global phase rather than extra driving fields to significantly suppress the systematic errors on LD and from quasiparticle poisoning. We also propose a protocol to realize the perfect chiral population transfer among three MZMs through their direct coupling with LD. An effective Hamiltonian can be constructed for the excited state of the local defect and the three MZMs, by which the population can propagate in both clockwise and counterclockwise manners among three MZMs with a unit fidelity. Our protocol demonstrates both braiding operations and chiral population transfer without a time-reversal-symmetry broken Hamiltonian, which are fundamental and essential elements for the topological quantum computation and quantum information processing.

It is clear to observe the discrepancies between the existing control protocols for Majorana zero modes and ours. The adiabatic braiding operations can be realized through projective measurements~\cite{Knapp2016Nature} and Floquet engineering~\cite{Bauer2019Topologically}. Although a near-unit fidelity is always achievable under ideal situations, the infidelity for both Floquet-engineering and measurement-based braiding operations inversely scales with the square of control period, i.e., $(1-\mathcal{F})\sim\mathcal{O}(T^{-2})$. The measurement-based braiding operation seems faster than the Floquet-engineering protocol since the effective coupling strength between MZMs of the former is about one order higher than that of the latter. However, it demands multiple projective measurements to perform the braiding operation, and then entails a significant overhead due to the probabilistic outcome. Our protocol takes advantage in speed since in principle all the nonadiabatic passages for constructing braiding operations can be implemented within a short period constrained only by the system characteristic frequency. While all the adiabatic control is much slower than the time-variation of the system parameters. The overhead of our protocol lies mainly in the control over the hopping and pairing couplings between the mediator LD and the neighboring fermions in the Kitaev chains and the correction operations performed to suppress the systematic errors on LD. The shortcut-to-adiabaticity version of braiding operation~\cite{Karzig2015Shortcuts} requires direct and time-dependent exchange interactions among MZMs, which is beyond the scope of the current technique. A recent nonadiabatic braiding operation~\cite{Yu2025Nonadiabatic} is based on the parallel-transport condition and is therefore fragile under the systematic errors. In comparison to both adiabatic protocols and existing nonadiabatic protocols on MZMs, our protocol only exploits the time-dependence of the local driving fields on an ancillary two-level system (local lattice defect), does not involve direct manipulation over MZMs, and is widely adaptable to other control tasks, such as chiral transfer.

\section*{Acknowledgments}

We acknowledge grant support from the ``Pioneer'' and ``Leading Goose'' R\&D of Zhejiang Province (Grant No. 2025C01028).

\appendix

\section{UQC theory in parametric space}\label{UniPara}

This appendix extends the universal quantum control theory~\cite{Jin2025Universal,Jin2025Entangling,Jin2025ErrCorr,Jin2025Rydberg} from the time domain to the parametric space. Consider a $K$-dimensional system under a general effective Hamiltonian $H_{\rm eff}(\vec{R})$ with a varying control parameter-vector $\vec{R}\equiv(R_1, R_2, \cdots, R_L)$, exhibiting an explicit or implicit time dependence, i.e., $R_j\equiv R_j(t)$ with $1\leq j\leq L$. $\dot{\vec{R}}\equiv d\vec{R}/dt\approx0$ means the adiabatic limit, by which the time evolution of the system is driven out of equilibrium at an infinitely slow rate. Our theory is however nonadiabatic. $H_{\rm eff}(\vec{R})$ can be obtained by such as Eq.~(\ref{Heff}). The system dynamics is described by the parameter-dependent Schr\"odinger equation
\begin{equation}\label{SchEff}
i\frac{\partial|\psi_n(\vec{R})\rangle}{\partial t}=H_{\rm eff}(\vec{R})|\psi_n(\vec{R})\rangle,
\end{equation}
where $|\psi_n(\vec{R})\rangle$'s are the pure-state solutions, $1\leq n\leq K$, constituting a completed and orthonormal set of the interested system.

In UQC theory~\cite{Jin2025Universal,Jin2025Entangling,Jin2025ErrCorr}, it is convenient to treat the system dynamics within an ancillary picture spanned by a completed and orthonormal set of ancillary basis states, i.e., $|\mu_k(\vec{R})\rangle$'s, $1\leq k\leq K$. $|\mu_k(\vec{R})\rangle$'s span the same Hilbert space as $|\psi_n(\vec{R})\rangle$'s. In the rotating frame with respect to $V(\vec{R})\equiv\sum_{k=1}^{K}|\mu_k(\vec{R})\rangle\langle\mu_k(\vec{R}_0)|$, where $\vec{R}_0$ is a fixed parameter vector representing but not limited to some boundary condition $\vec{R}(t=t_0)$, the Hamiltonian can be transformed as
\begin{equation}\label{Hamrot}
\begin{aligned}
&H_{\rm rot}(\vec{R})=V^\dagger(\vec{R})H_{\rm eff}(\vec{R})V(\vec{R})-iV^\dagger(\vec{R})\partial_tV(\vec{R})\\
&=\sum_{k=1}^{K}\sum_{n=1}^{K}\Big[\langle\mu_{k}(\vec{R})|H_{\rm eff}(\vec{R})|\mu_n(\vec{R})\rangle\\
&-i\sum_{j=1}^M\dot{R}_j\langle\mu_k(\vec{R})|\partial_{R_j}\mu_n(\vec{R})\rangle\Big]|\mu_k(\vec{R}_0)\rangle\langle\mu_n(\vec{R}_0)|\\
&\equiv\sum_{k=1}^{K}\sum_{n=1}^{K}\left[\mathcal{H}_{kn}(\vec{R})-\mathcal{A}_{kn}(\vec{R})\right]|\mu_k(\vec{R}_0)\rangle\langle\mu_n(\vec{R}_0)|\\
&=V^\dagger(\vec{R})\left[\mathcal{H}(\vec{R})-\mathcal{A}(\vec{R})\right]V(\vec{R}),
\end{aligned}
\end{equation}
where the dynamical component $\mathcal{H}_{kn}(\vec{R})$ and the gauge-potential component $\mathcal{A}_{kn}(\vec{R})$~\cite{Michael2017Geometry,Claeys2019Floquet,Takahashi2024Shortcuts} constitute the elements of the matrices $\mathcal{H}(\vec{R})$ and $\mathcal{A}(\vec{R})$, respectively, at the $k$th row and the $n$th column. The purely geometrical matrix $\mathcal{A}(\vec{R})$ relates uniquely to the manifold geometry determined by the ancillary basis states $|\mu_k(\vec{R})\rangle$'s, whose Berry connection is $i\langle\mu_k(\vec{R})|\nabla_{\vec{R}}|\mu_n(\vec{R})\rangle$. The Schr\"odinger equation~(\ref{SchEff}) can then be transformed as
\begin{equation}\label{Schrot}
i\frac{\partial|\psi_n(\vec{R})\rangle_{\rm rot}}{\partial t}=H_{\rm rot}(\vec{R})|\psi_n(\vec{R})\rangle_{\rm rot}
\end{equation}
with the rotated pure-state solutions $|\psi_n(\vec{R})\rangle_{\rm rot}$
\begin{equation}\label{relate}
|\psi_n(\vec{R})\rangle_{\rm rot}\equiv V^\dagger(\vec{R})|\psi_n(\vec{R})\rangle.
\end{equation}

The parameter-dependent Schr\"odinger equation (\ref{Schrot}) can be exactly solved when $H_{\rm rot}(\vec{R})$ is diagonalizable~\cite{Jin2025Universal,Jin2025Entangling,Jin2025ErrCorr}, or more precisely, when Eq.~(\ref{Hamrot}) can be simplified as
\begin{equation}\label{Hdigfull}
H_{\rm rot}(\vec{R})=\sum_{k=1}^{K}\left[\mathcal{H}_{kk}(\vec{R})-\mathcal{A}_{kk}(\vec{R})\right]
|\mu_k(\vec{R}_0)\rangle\langle\mu_k(\vec{R}_0)|.
\end{equation}
In this case, the evolution operator $U_{\rm rot}(\vec{R})$ can be directly obtained by Eq.~(\ref{Hdigfull}) as
\begin{equation}\label{Urot1}
U_{\rm rot}(\vec{R})=\sum_{k=1}^{K}e^{if_k(\vec{R})}|\mu_k(\vec{R}_0)\rangle\langle\mu_k(\vec{R}_0)|,
\end{equation}
where the global phases $f_k(\vec{R})$'s are defined as
\begin{equation}\label{global}
f_k(\vec{R})\equiv\int_{\vec{R}_0}^{\vec{R}}\frac{1}{\dot{\vec{R}}'}
\left[\mathcal{A}_{kk}(\vec{R}')-\mathcal{H}_{kk}(\vec{R}')\right]d\vec{R}'.
\end{equation}
Transformed back to the original picture via Eq.~(\ref{relate}), the evolution operator can be written as
\begin{equation}\label{U0}
U_0(\vec{R})=V(\vec{R})U_{\rm rot}(\vec{R})=\sum_{k=1}^{K}e^{if_k(\vec{R})}|\mu_k(\vec{R})\rangle\langle\mu_k(\vec{R}_0)|.
\end{equation}

Due to the theory of UQC, the diagonalization condition for $H_{\rm rot}(\vec{R})$~(\ref{Hamrot}) in the $k$th ancillary basis state $|\mu_k(\vec{R})\rangle$ is given by
\begin{equation}\label{ConditionDiag}
\left[H_{\rm eff}(\vec{R})-\mathcal{A}(\vec{R}), \Pi_k(\vec{R})\right]=0
\end{equation}
with the projector $\Pi_k(\vec{R})\equiv|\mu_k(\vec{R})\rangle\langle\mu_k(\vec{R})|$. It admits a compact form as
\begin{equation}\label{Diagdedu}
\begin{aligned}
&\left[H_{\rm eff}(\vec{R}), \Pi_k(\vec{R})\right]=\left[\mathcal{A}(\vec{R}), \Pi_k(\vec{R})\right]\\
=&i\sum_{j=1}^M\dot{R}_j\left[|\partial_{R_j}\mu_k(\vec{R})\rangle\langle\mu_k(\vec{R})|+
|\mu_k(\vec{R})\rangle\langle\partial_{R_j}\mu_k(\vec{R})|\right]\\
=&i\dot{\vec{R}}\cdot\left[|\nabla_{\vec{R}}\mu_k(\vec{R})\rangle\langle\mu_k(\vec{R})|
+|\mu_k(\vec{R})\rangle\langle\nabla_{\vec{R}}\mu_k{(\vec{R})}|\right]\\
=&i\dot{\vec{R}}\cdot\nabla_{\vec{R}}\Pi_k(\vec{R}),
\end{aligned}
\end{equation}
where the second equivalence used the relation for the gauge field, i.e., $\mathcal{A}(\vec{R})|\mu_k(\vec{R})\rangle=i\sum_{j=1}^M\dot{R}_j|\partial_{R_j}\mu_k(\vec{R})\rangle$, as defined in Eq.~(\ref{Hamrot}). Equation~(\ref{Diagdedu}) can be recast as a von Neumann-type equation~\cite{Jin2025Universal,Jin2025Entangling,Jin2025ErrCorr}:
\begin{equation}\label{von}
\frac{\partial}{\partial t}\Pi_k(\vec{R})=-i\left[H_{\rm eff}(\vec{R}), \Pi_k(\vec{R})\right].
\end{equation}
When $k$ in Eq.~(\ref{von}) runs from 1 to $K$, $H_{\rm rot}(\vec{R})$ can be completely diagonalized.

\section{UQC theory in parametric space with error correction}\label{UniParaErr}

This appendix provides a recipe for the universal quantum control theory in parametric space to suppress the systematic errors~\cite{Jin2025ErrCorr}. When the control parameters are subject to unavoidable fluctuations or unwanted crosstalk with the external degrees of freedom, the system dynamics is actually described by the realistic Hamiltonian $H(\vec{R})=H_{\rm eff}(\vec{R})+\epsilon H_1(\vec{R})$, where $H_1(\vec{R})$ is the error Hamiltonian and $\epsilon$ is a perturbative coefficient. Therefore, the parameter-dependent Schr\"odinger equation~(\ref{SchEff}) becomes
\begin{equation}\label{SchEfferr}
i\frac{\partial|\psi_n(\vec{R})\rangle}{\partial t}=H(\vec{R})|\psi_n(\vec{R})\rangle.
\end{equation}

Consider the system dynamics in the rotating frame with respect to $V(\vec{R})\equiv\sum_{k=1}^{K}|\mu_k(\vec{R})\rangle\langle\mu_k(\vec{R}_0)|$, the ideal rotated Hamiltonian in Eq.~(\ref{Hdigfull}) can be perturbed as
\begin{equation}\label{HamrotNon}
\begin{aligned}
H_{\rm rot}(\vec{R})&=-\sum_{k=1}^{K}\dot{\vec{R}}\cdot\nabla_{\vec{R}}f_k(\vec{R})|\mu_k(\vec{R}_0)\rangle\langle\mu_k(\vec{R}_0)|\\
&+\epsilon\sum_{k=1}^{K}\sum_{n=1}^{K}\mathcal{D}_{kn}^{\rm err}(\vec{R})|\mu_k(\vec{R}_0)\rangle\langle\mu_n(\vec{R}_0)|,
\end{aligned}
\end{equation}
where $\mathcal{D}_{kn}^{\rm err}(\vec{R})\equiv\langle\mu_{k}(\vec{R})|H_1(\vec{R})|\mu_n(\vec{R})\rangle$. Equation~(\ref{HamrotNon}) indicates that the errors can yield unwanted transitions of the system from the passage $|\mu_k(\vec{R})\rangle$ to the passage $|\mu_n(\vec{R})\rangle$, $n\neq k$.

To identify and neutralize the adverse effects induced by systematic errors, the system dynamics can be analyzed in the double-rotated picture with respect to $U_{\rm rot}(\vec{R})$ in Eq.~(\ref{Urot1}). In particular, we have
\begin{equation}\label{HamNonSecRot}
\begin{aligned}
H_{\rm err}(\vec{R})&=U_{\rm rot}^\dagger(\vec{R})H_{\rm rot}(\vec{R})U_{\rm rot}(\vec{R})-iU_{\rm rot}^\dagger(\vec{R})\partial_tU_{\rm rot}(\vec{R})\\
&=\epsilon\sum_{k=1}^{K}\sum_{n=1}^{K}\tilde{\mathcal{D}}_{kn}^{\rm err}(\vec{R})|\mu_k(\vec{R}_0)\rangle\langle\mu_n(\vec{R}_0)|
\end{aligned}
\end{equation}
where
\begin{equation}\label{ErrElement}
\tilde{\mathcal{D}}_{kn}^{\rm err}(\vec{R})\equiv\langle\mu_k(\vec{R})|H_1(\vec{R})|\mu_n(\vec{R})\rangle e^{-i[f_k(\vec{R})-f_n(\vec{R})]}.
\end{equation}

Using the Magnus expansion~\cite{Blanes2009Magnus}, the evolution operator for $H_{\rm err}(\vec{R})$ in Eq.~(\ref{HamNonSecRot}) can be expanded as
\begin{equation}\label{Uerr}
\begin{aligned}
&U_{\rm err}(\vec{R})\\
&=1-i\int_{\vec{R}_0}^{\vec{R}}d\vec{R}'\frac{H_{\rm err}(\vec{R}')}{\dot{\vec{R}}'}-\frac{1}{2}\Big\{\left[\int_{\vec{R}_0}^{\vec{R}}d\vec{R}'\frac{H_{\rm err}(\vec{R}')}{\dot{\vec{R}}'}\right]^2\\
&+\int_{\vec{R}_0}^{\vec{R}}d\vec{R}'\left[\frac{H_{\rm err}(\vec{R}')}{\dot{\vec{R}}'}, \int_{\vec{R}_0}^{\vec{R}'}d\vec{R}''\frac{H_{\rm err}(\vec{R}'')}{\dot{\vec{R}}''}\right]\Big\}+\cdots.
\end{aligned}
\end{equation}
To evaluate the impact of the systematic errors, one can define the overlap or fidelity between the instantaneous state governed by $U_{\rm err}(\vec{R})$ in Eq.~(\ref{Uerr}) and the target passage $|\mu_k(\vec{R}_0)\rangle$ in the rotating frame~\cite{Jin2025ErrCorr} as
\begin{equation}\label{fidelity}
\begin{aligned}
\mathcal{F}&\equiv\left|\langle\mu_k(\vec{R}_0)|U_{\rm err}(\vec{R})|\mu_k(\vec{R}_0)\rangle\right|^2\\
&\approx1-\epsilon^2\sum_{n=1,n\ne k}^K\left|\int_{\vec{R}_0}^{\vec{R}}\frac{\tilde{\mathcal{D}}_{kn}^{\rm err}(\vec{R}')}{\dot{\vec{R}}'}d\vec{R}'\right|^2+\mathcal{O}(\epsilon^3).
\end{aligned}
\end{equation}
The error up to the second order of $\epsilon$ can be canceled under the error correction mechanism~\cite{Jin2025ErrCorr}
\begin{equation}\label{OptGeneral}
\left|\nabla_{\vec{R}}\left[f_k(\vec{R})-f_n(\vec{R})\right]\right|\gg\left|\nabla_{\vec{R}}\left[
\langle\mu_k(\vec{R})|H_1(\vec{R})|\mu_n(\vec{R})\rangle\right]\right|.
\end{equation}
This result is supported by the nonperturbative dynamical decoupling method~\cite{Jing2013Nonperturbative,Jing2015Nonperturbative} and the Riemann-Lebesgue lemma in mathematics.

When the control parameter vector $\vec{R}$ only involves one variable $R$, i.e., $\vec{R}\rightarrow R$, Eq.~(\ref{OptGeneral}) can be reduced to
\begin{equation}\label{OptGeneralOne}
\left|\frac{d}{dR}[f_k(R)-f_n(R)]\right|\gg\left|\frac{d\langle\mu_k(R)|H_1(R)|\mu_n(R)\rangle}{dR}\right|.
\end{equation}
Also, the gradient inequality with respect to the control parameter can be reformulated as a gradient inequality with respect to time:
\begin{equation}\label{OptGeneraltIime}
\left|\frac{d}{dt}[f_k(t)-f_n(t)]\right|\gg\frac{d}{dt}\left[\langle\mu_k(t)|H_1(t)|\mu_n(t)\rangle\right],
\end{equation}
which has been proposed in Ref.~\cite{Jin2025ErrCorr}.

\bibliographystyle{apsrevlong}
\bibliography{ref}

\end{document}